\documentclass[Physsubmission, Phys]{SciPost}

\usepackage[utf8]{inputenc}
\usepackage{tikz}

\graphicspath{{figures/}}

\hypersetup{
    colorlinks,
    linkcolor={red!50!black},
    citecolor={blue!50!black},
    urlcolor={blue!80!black}
}

\usepackage[bitstream-charter]{mathdesign}
\DeclareSymbolFont{usualmathcal}{OMS}{cmsy}{m}{n}
\DeclareSymbolFontAlphabet{\mathcal}{usualmathcal}
\begin{document}

\begin{center}{\Large \textbf{
Forward Proton Measurements with ATLAS}
}
\end{center}

\begin{center}
M. Schmidt\textsuperscript{1$\star$} on behalf of the ATLAS Collaboration\footnote{Copyright 2022 CERN for the benefit of the ATLAS Collaboration. Reproduction of this article or parts of it is allowed as specified in the CC-BY-4.0 license.}
\end{center}

\begin{center}
{\bf 1} University of Wuppertal, Germany
\\
*muschmidt@uni-wuppertal.de
\end{center}

\begin{center}
\today
\end{center}

\definecolor{palegray}{gray}{0.95}
\begin{center}
\colorbox{palegray}{
  \begin{tabular}{rr}
  \begin{minipage}{0.1\textwidth}
    \includegraphics[width=23mm]{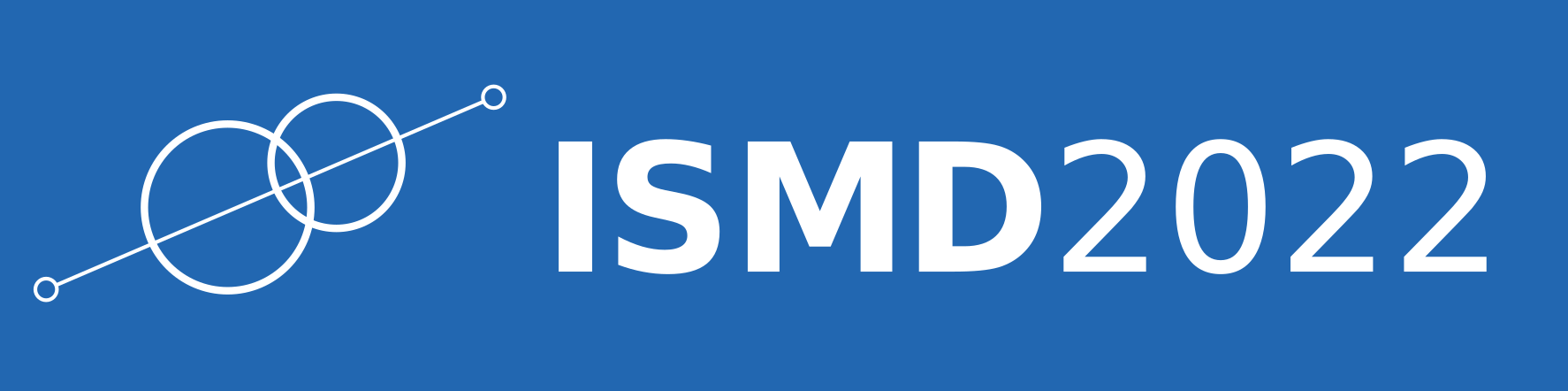}
  \end{minipage}
  &
  \begin{minipage}{0.8\textwidth}
    \begin{center}
    {\it 51st International Symposium on Multiparticle Dynamics (ISMD2022)}\\ 
    {\it Pitlochry, Scottish Highlands, 1-5 August 2022} \\
    \doi{10.21468/SciPostPhysProc.?}\\
    \end{center}
  \end{minipage}
\end{tabular}
}
\end{center}

\section*{Abstract}
{\bf
The ALFA subdetector is designed to measure elastic proton-proton scattering from which the total cross section and $\rho$-parameter are determined.
In 2016, special runs at a center-of-mass energy of $\sqrt{s} = 13$\,TeV and with $\beta^\star = 2.5$\,km were recorded.
}

\vspace{10pt}
\noindent\rule{\textwidth}{1pt}
\tableofcontents\thispagestyle{fancy}
\noindent\rule{\textwidth}{1pt}
\vspace{10pt}

\section{Introduction}
\begin{figure}
    \centering
    \includegraphics[width=0.4\textwidth]{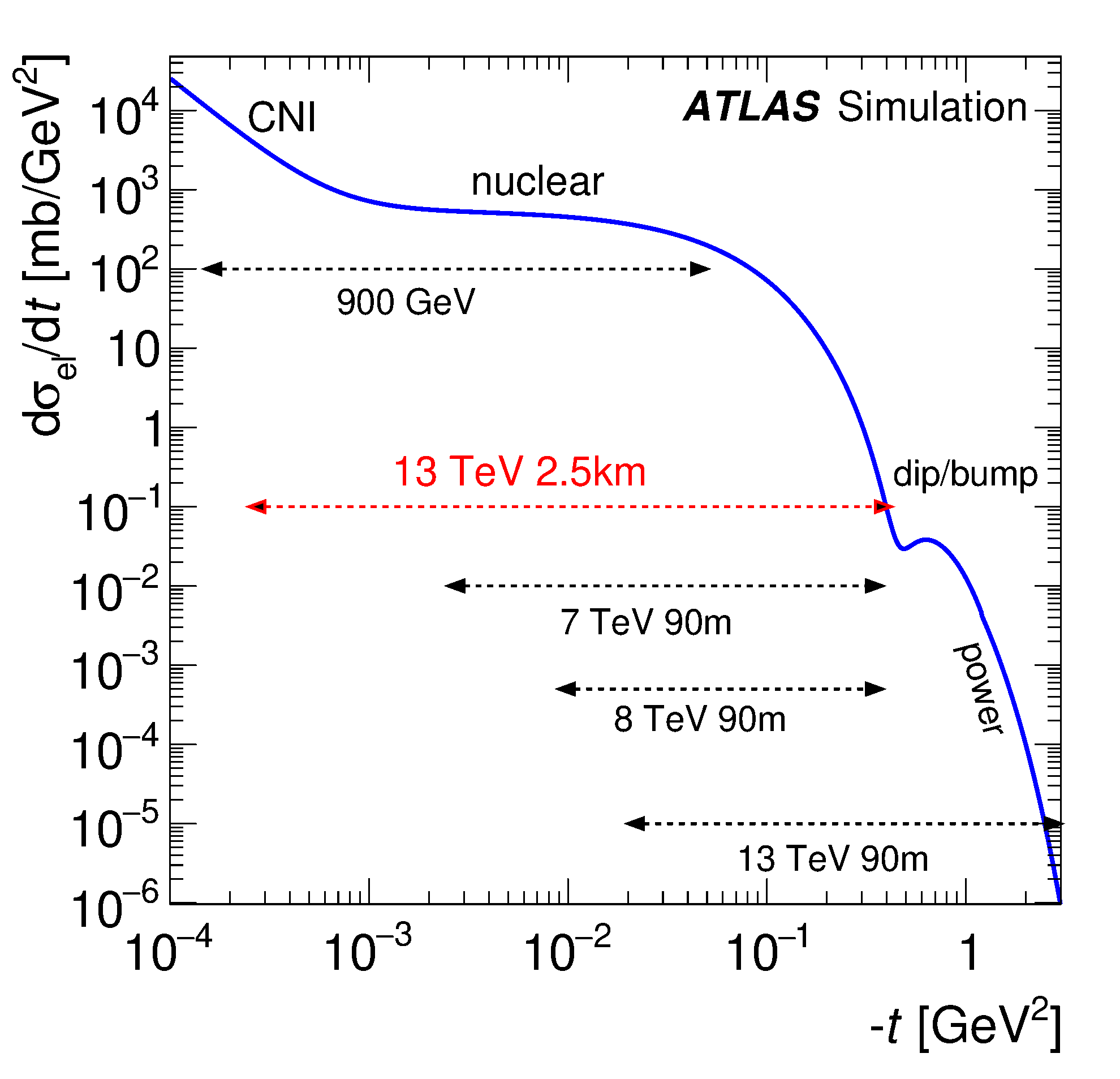}
    \caption{Model prediction of the differential elastic cross-section \cite{briefing}}
    \label{fig:prediction}
\end{figure}

The goal of this measurement campaign \cite{2022mgx} is to study elastic proton-proton scattering, especially within the very interesting Coulomb-Nuclear-Interference (CNI) region.
In order to study the energy evolution, for these runs special optics with large $\beta^\star$ values of 2.5\,km in combination with high center-of-mass energy of $\sqrt{s}=13$\,TeV have been chosen which makes it possible to reach down to very small momentum-transfers as illustrated in Figure~\ref{fig:prediction}.

An important quantity to be measured is the $\rho$-parameter which is defined as the fraction of the real part of the elastic scattering amplitude and the imaginary part for small $t$:
\begin{equation}
    \rho = \left.\frac{\mathrm{Re}[f_\mathrm{el}(t)]}{\mathrm{Im}[f_\mathrm{el}(t)]}\right|_{t\rightarrow 0}
\end{equation}
The total cross-section can be calculated from the differential elastic cross-section with the help of the optical theorem:
\begin{equation}
    \sigma_\mathrm{tot} = 4\pi\,\mathrm{Im}\,[f_\mathrm{el} (t\rightarrow 0)]
\end{equation}
The description of this analysis is divided into 2 parts.
This document covers the experimental setup and theoretical background.
\section{Experimental Setup}
\begin{figure}[h]
\begin{center}
\includegraphics[width=0.45\linewidth]{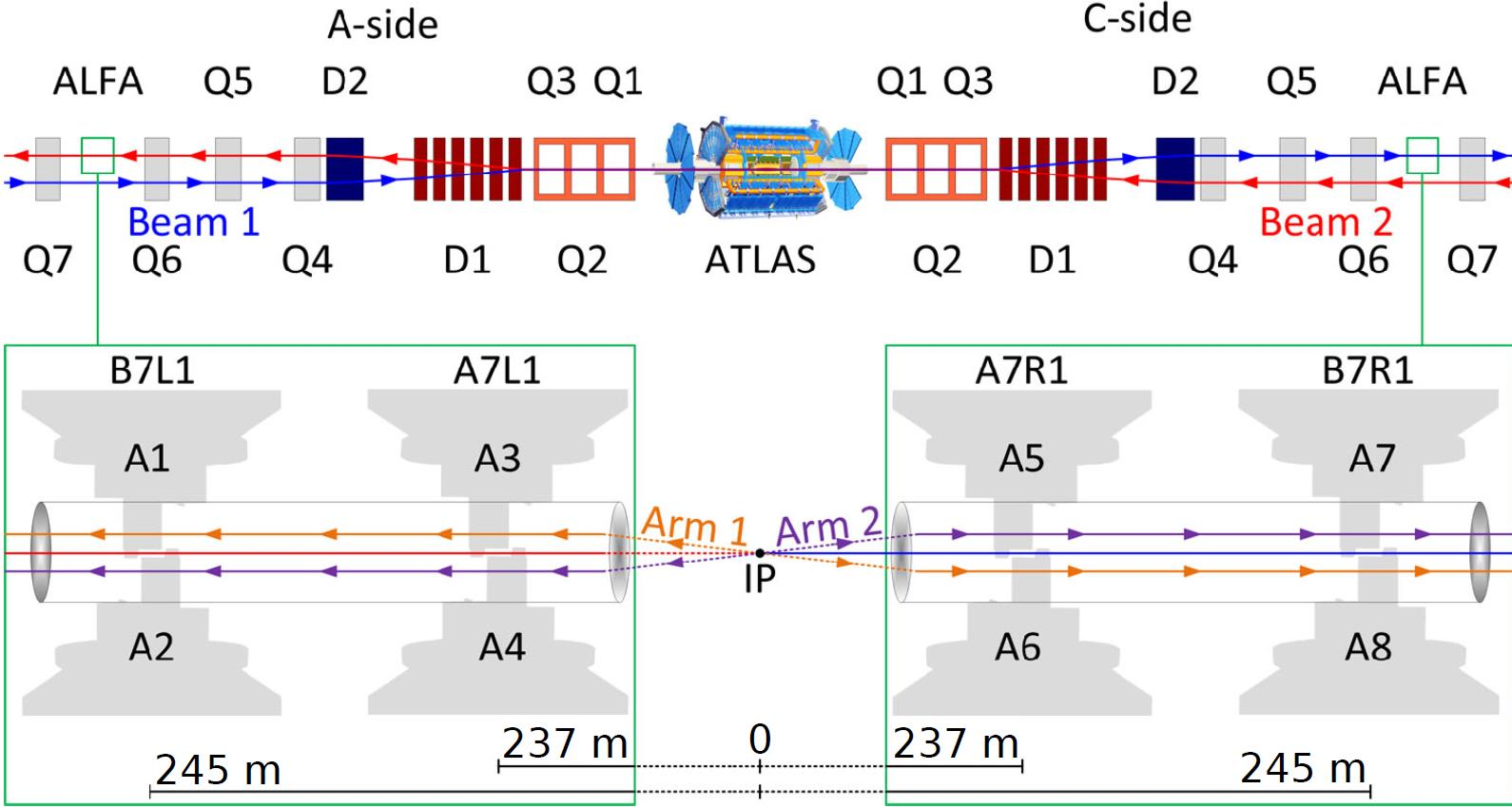}
\includegraphics[width=0.4\linewidth]{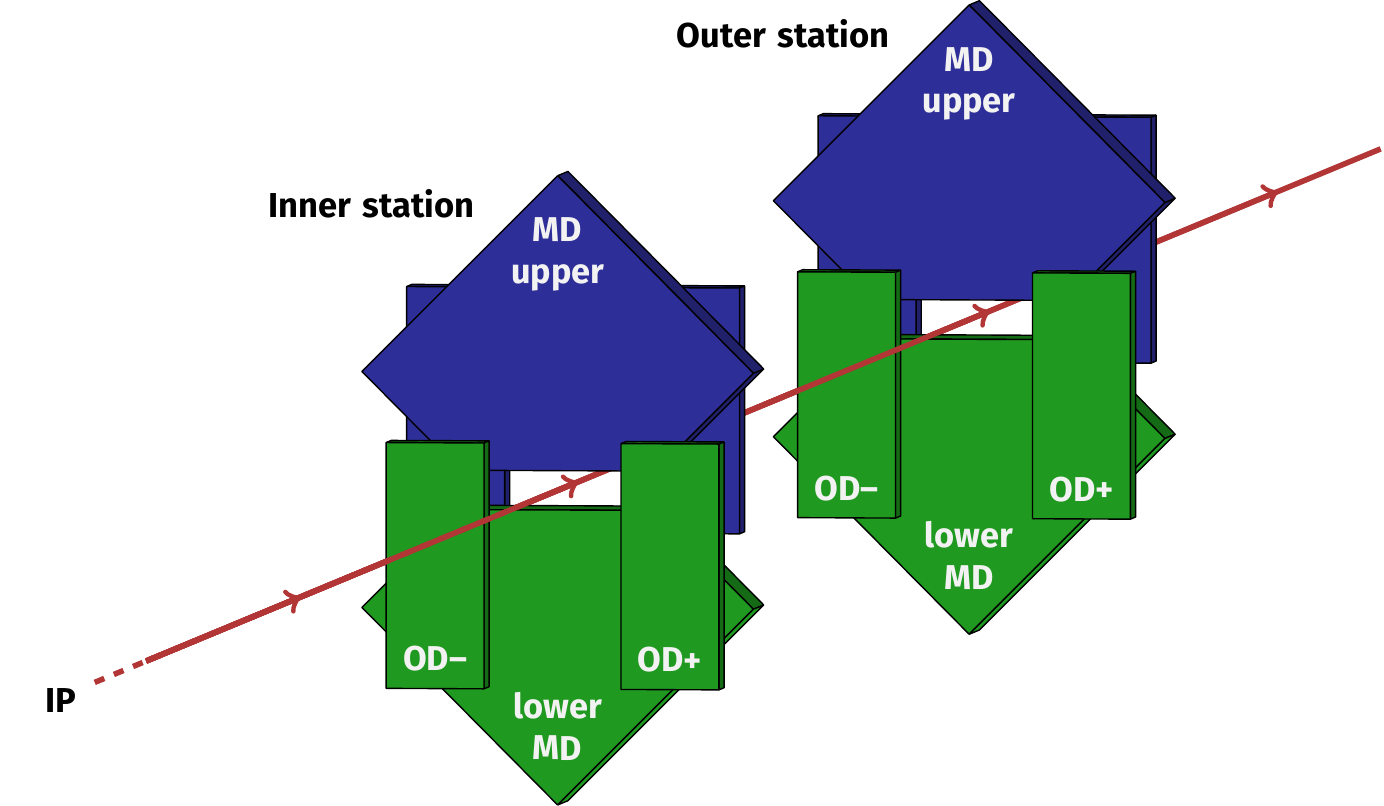}
\caption{Experimental Setup of ALFA and a sketch of 2 roman pot stations on one side of the detector in the outgoing beam direction \cite{2022mgx}}
\label{fig:setup}
\end{center}
\end{figure}

The ALFA detector \cite{ALFA} contains 4 roman pot (RP) detector stations that are placed around 240\,m away from the interaction point (IP) of the ATLAS detector.
The inner stations on the A- and C-side have a distance of 237\,m to the IP,  whereas the outer stations are positioned 245\,m from the IP. 
A sketch of the experimental setup on the left side of  Figure~\ref{fig:setup} illustrates the arrangement of all stations and detectors.

Due to the back-to-back scattering of both elastically interacting protons, the 8 detectors can be divided into 2 groups.
The upper detectors at side A and the lower detectors at side C are combined to Arm 1, whereas Arm 2 contains the other 4 detectors.
Each station contains 2 main detectors (upper and lower MDs) and 2 overlap detectors (OD+ and OD-) that contain an upper and a lower part.
The MDs contain $2\times 10$ layers of 64 scintillating fibers with a side length of 0.5\,mm attached to 64-channel Multi-Anode Photomultipliers (MaPMTs).
If a proton hits these fibers, the produced photons are internally reflected in the fibers and kick out a photoelectron when reaching the photocathode of the related MaPMT.
The resulting avalanche is then detected on the segmented anode.
For digitization, an analog Multi-Anode ReadOut Chip (MAROC) in combination with a Field-Programmable Gate Array (FPGA) is used.
From the channel number related to the hit fiber, the proton position in each layer can be reconstructed and converted into $x$ and $y$-coordinates.

The right side of Figure~\ref{fig:setup} shows a sketch of 2 RP stations including both MDs and ODs on the A- or C-side in the beam direction of the outgoing protons from the IP.
The upper and lower parts are colored in blue and green respectively.

\section{Kinematics}
\begin{figure}[h]
\begin{center}
    \begin{tikzpicture}[scale=0.7]
        \draw[red,-latex] (0,0) -- ++(30:3)  node[right] {$p_3$};
        \draw[red,-latex] (0,0) -- ++(-150:3)  node[below] {$p_4$};
        \draw[-latex] (-3,0) node[left] {$p_1$} -- ++(3,0);
        \draw[latex-] (0,0) -- ++(3,0) node[right] {$p_2$};
        \fill[black!50] (0,0) circle (2mm);
        \draw (0,0) ++(0:2) arc (0:30:2);
        \draw (15:1.5) node {$\theta$};

        \draw[-latex] (5,1) -- (9,1) node[midway, above] {$p_1$};
        \draw[dashed] (9,1) -- (13,1) node (a){};
        \draw[-latex] (9,1) -- ++(-40:4) node (b){} node [midway, below] {$p_3$};
        \draw [latex-]  (a.center) -- (b.center) node[midway, right] {$q = p_3 - p_1$};
        \draw[dashed] (9,1) -- ++(-20:3.75);
        \draw (9,1) ++(0:2) arc (0:-20:2);
        \draw (9,1) ++(-20:2.5) arc (-20:-40:2.5);
        \draw (9,1) ++(-30:2) node {$\theta/2$};
    \end{tikzpicture}
\end{center}
\caption{Kinematics of elastic proton-proton scattering}
\label{fig:kinematics}
\end{figure}
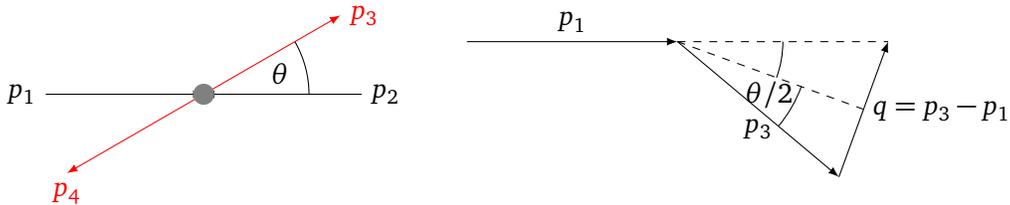

During the elastic scattering process of the 2 interacting protons, the total energy and momentum of both particles are conserved.
Hence, the momentum transfer $t$, which is defined as the square of the difference between the momenta of the incoming and outgoing proton, can be calculated according to
\begin{equation}
    t = q^2 = (p_1 - p_3)^2 = 4p^2\sin^2\left(\frac{\theta}{2}\right)
\end{equation}
as shown in Figure~\ref{fig:kinematics}.
For small scattering angles, the sine function can be approximated according to $\sin\theta\approx\theta$ which leads to the following relation:
\begin{equation}
    t\approx (p\theta)^2
\end{equation}
The momentum transfer is therefore simply given as the squared product of the beam momentum and the scattering angle.
Since the beam momentum is known, the scattering angle of every single event has to be determined with ALFA, in order to reconstruct the momentum transfer for every scattered proton pair.

\section{Reconstruction}
Before the $t$-spectrum can be reconstructed, an alignment procedure has to be performed which mainly consists of an iterative procedure and a $\chi^2$-optimization \cite{STDM-2013-10}.
For improving the results, a physics-driven alignment \cite{2022mgx} is additionally required for this analysis.
After the elastic scattering process, the outgoing protons are deflected by LHC quadrupole magnets.
The measured angle $\theta$ is therefore not identical to the scattering angle $\theta^\star$ at the IP.
In order to calculate $\theta^\star$, the beam optics transport matrix elements have to be taken into account.
Several reconstruction methods with different levels of precision have been developed throughout the past years.
For the $t$-spectrum reconstruction of this measurement campaign, the so-called subtraction method delivers the best performance.

This method contains the computation of the difference between the measured $x$ and $y$-coordinate of each detector side.
The result then has to be divided by the sum of the matrix elements $M_{12}$ for both sides which directly results in $\theta^\star$:
\begin{equation}
    \theta^* = \frac{\{x,y\}_A-\{x,y\}_C}{M_{12,A} + M_{12,C}}
\end{equation}

\begin{figure}
    \centering
    \includegraphics[width=0.4\textwidth]{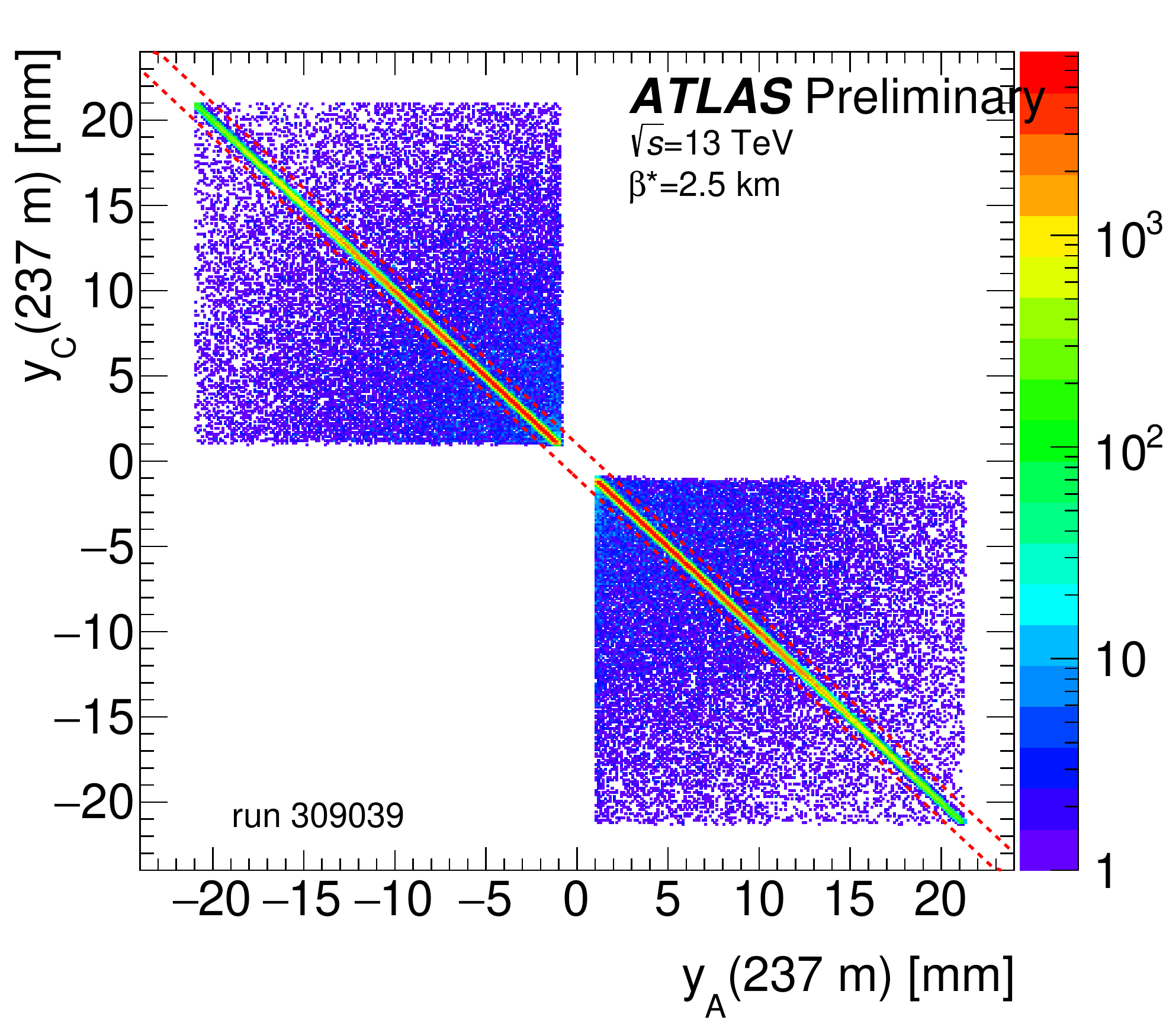}
    \caption{Selecting elastic events from correlation between $y$-values \cite{2022mgx}}
    \label{fig:selection}
\end{figure}

With the help of the determined momentum transfer of each event and the number of events within the covered $t$-range, the elastic cross-section can be calculated.
The full analysis contains several steps such as the selection of elastic events as shown in Figure~\ref{fig:selection} for the example of the correlation between $y$-values of both detector sides.
Additionally, background subtraction, acceptance, unfolding corrections, reconstruction efficiency, and normalization with the luminosity, measured independently, are required.

The total cross-section is then given by
\begin{equation}
    \frac{\mathrm{d}\sigma}{\mathrm{d}t} = \frac{1}{16\pi}\left|-8\pi\alpha\hbar c \frac{G^2(t)}{|t|} e^{i\alpha \phi(t)} + (\rho +i)\frac{\sigma_\mathrm{tot}}{\hbar c}e^{\frac{-B|t|-Ct^2-D|t|^3}{2}}\right|^2
\end{equation}
The left addend represents the Coulomb amplitude, whereas the right one stands for the nuclear term.
Therefore, $G$ denotes the electric form factor of the proton and $B$ the nuclear slope parameter.
This formula is only valid for small momentum transfers $t\rightarrow 0$.
All parameters including $\rho$ and $\sigma_\mathrm{tot}$ can be determined with a fit to the reconstructed $t$-spectrum.
The inelastic cross-section is then simply given by the difference between the total cross-section and total elastic cross-section:
\begin{equation}
\sigma_\mathrm{inel} = \sigma_\mathrm{tot} - \sigma_\mathrm{el}
\end{equation}
The total elastic cross-section is determined by integrating the nuclear part of the differential elastic cross-section. The inelastic cross-section is therefore not an independent measurement, but rather a derived quantity based on the main fit result.
\bibliography{references.bib}
\end{document}